\documentclass[aps,pra,twocolumn,amsmath,amssymb]{revtex4-1}

\usepackage{graphicx}
\usepackage[utf8]{inputenc}
\usepackage{textcomp}
\usepackage{mathtools}
\usepackage{hyperref}
\usepackage[all]{hypcap}
\input{glyphtounicode}
\begin{document}

\makeatletter
\renewcommand{\fnum@figure}{\textbf{Figure~\thefigure}}
\makeatother

\title{Frequency splitting of polarization eigenmodes in microscopic Fabry-Perot cavities}

\author{Manuel Uphoff}
\author{Manuel Brekenfeld}
\author{Gerhard Rempe}
\author{Stephan Ritter}
\email{stephan.ritter@mpq.mpg.de}
\affiliation{Max-Planck-Institut für Quantenoptik, Hans-Kopfermann-Strasse 1, 85748 Garching, Germany}

\begin{abstract}
We study the frequency splitting of the polarization eigenmodes of the fundamental transverse mode in CO$_2$ laser-machined, high-finesse optical Fabry-Perot cavities and investigate the influence of the geometry of the cavity mirrors. Their highly reflective surfaces are typically not rotationally symmetric but have slightly different radii of curvature along two principal axes. We observe that the eccentricity of such elliptical mirrors lifts the degeneracy of the polarization eigenmodes. The impact of the eccentricity increases for smaller radii of curvature. A model derived from corrections to the paraxial resonator theory is in excellent agreement with the measurements, showing that geometric effects are the main source of the frequency splitting of polarization modes for the type of microscopic cavity studied here. By rotating one of the mirrors around the cavity axis, the splitting can be tuned. In the case of an identical differential phase shift per mirror, it can even be eliminated, despite a nonvanishing eccentricity of each mirror. We expect our results to have important implications for many experiments in cavity quantum electrodynamics, where Fabry-Perot cavities with small mode volumes are required.
\end{abstract}

\maketitle

\section{Introduction}
Fabry-Perot resonators entered the scene of physics more than a century ago and have continued to play an important role ever since. A very prominent entry was the advent of the laser, where a Fabry-Perot cavity provides the optical feedback required for the coherent amplification of light. Since then, many efforts have been targeted at increasing the finesse and decreasing the mode volume of these resonators in order to enhance the coupling of light to progressively smaller amounts of matter. Advances in these directions have resulted in rapidly developing fields like cavity quantum electrodynamics (CQED) and cavity optomechanics \cite{Aspelmeyer2014}. By now, research on CQED has reached the regime of single quanta of light and matter, as exemplified by the nondestructive detection of a photon \cite{Nogues1999,Reiserer2013} or a universal single-atom quantum network node \cite{Ritter2012,Stute2013,Reiserer2014}.

The quest for even smaller mode volumes with dimensions approaching the wavelength has triggered the development of new mirror fabrication techniques for Fabry-Perot resonators \cite{Trupke2005,Steinmetz2006,Dolan2010,Hunger2010,Muller2010} and led to numerous efforts towards new types of monolithic microresonators \cite{Aoki2006,Junge2013,Thompson2013}. In monolithic systems, atoms are coupled to the evanescent field of a tightly confined mode. In contrast, Fabry-Perot cavities offer polarization control and the advantage of easy access to the field mode \cite{Brennecke2007,Colombe2007,Herskind2009,Wolke2012,Chen2013}. Microfabricated Fabry-Perot cavities have been employed in a wide variety of contexts: they have yielded unprecedentedly high light-atom coupling rates in Fabry-Perot resonators \cite{Gehr2010b}, have been used to approach the strong-coupling regime of CQED with ions \cite{Steiner2013}, and have enabled CQED experiments with solid-state emitters in Fabry-Perot cavities \cite{Flagg2010,Di2012,Miguel-Sanchez2013,Albrecht2013,Kaupp2013}. They have further been utilized to enhance Raman scattering of molecules \cite{Petrak2014} and to couple light to micromechanical objects \cite{Shkarin2014}. The highest surface qualities for micromirrors have been achieved with a fabrication process using a CO$_2$ laser to shape the end facets of optical fibres \cite{Hunger2010,Muller2010,Hunger2012}. These are subsequently coated with Bragg mirrors, applying the same process routinely used for superpolished substrates.

To exploit the full potential of high-finesse Fabry-Perot cavities, the control of polarization eigenmodes is essential. Examples range from cavity-enhanced polarimetry \cite{DellaValle2013,Cadene2014,Durand2010} and cavity ring-down spectroscopy \cite{Muller2002,Huang2008} to applications in quantum information processing, such as the efficient and coherent coupling of atomic states to the polarization of single photons \cite{Wilk2007}. The latter requires degeneracy of the polarization eigenmodes, which has been achieved for Fabry-Perot cavities built from superpolished mirror substrates \cite{Munstermann1999}. Microfabricated cavities, however, can have increased frequency splittings between polarization eigenmodes, as was first observed in CO$_2$ laser-machined resonators \cite{Gehr2010b,Hunger2010}. If the splitting is on the order of the linewidth of the cavity, there can be detrimental effects on all kinds of experiments \cite{Hall2000, Kuhr2007, Volz2011}. There are two strategies for dealing with the splitting in cavities: to minimize it until it becomes negligible, or to increase it such that the two polarization modes are well separated \cite{Kuhr2007}. In either case, it is necessary to understand and control this splitting.

Two potential sources of the splitting of polarization eigenmodes in a Fabry-Perot cavity can be distinguished. The first one is birefringence of the mirror materials, usually attributed to mechanical stress \cite{Hall2000,Brandstatter2013}. Combined with a finite penetration depth, this leads to a polarization-dependent phase shift upon reflection.
The second source is directly related to the cavity geometry. Its existence is not evident from the usual paraxial resonator theory, in which the cavity field and its resonances are described by a scalar mode function that is independent of the polarization. The paraxial theory does describe the polarization-independent splitting of higher-order transverse modes of equal order in a cavity with elliptical mirrors, but it cannot account for an additional splitting of each of these modes into a doublet via the polarization degree of freedom. Any splitting of the polarization modes thus has to originate from effects beyond the paraxial approximation. It has been shown that corrections to the paraxial theory predict a splitting of polarization eigenmodes of higher-order transverse modes for cavities with spherical mirrors \cite{Yu1983,Luk1986,Foster2009,Zeppenfeld2010}. However, due to the cylindrical symmetry assumed in these calculations, they result in degenerate fundamental transverse polarization eigenmodes. Because fundamental transverse modes are of greatest practical relevance in CQED experiments, they are the subject of this work.

In the following, we show that corrections to the paraxial approximation can explain a frequency splitting of the fundamental transverse mode in cavities with elliptical mirrors. We find good quantitative agreement between experimental data and a theoretically derived analytic relation between surface geometry and induced frequency splitting. This confirms that for mirrors machined with a CO$_2$ laser, their ellipticity is the dominant reason for the splitting of polarization modes. Therefore, the expected splitting can be predicted from the surface data of fabricated structures, even before the application of a mirror coating. The studied effect is very general, and by no means exclusive to CO$_2$ laser-machined cavities, but appears whenever the radii of curvature of elliptical cavity mirrors approach the wavelength of the resonant field. Finally, we demonstrate that for cavities with two elliptical mirrors, the amount of frequency splitting can be further controlled by rotating one of the mirrors.

\begin{figure}
\includegraphics[width=0.9\columnwidth]{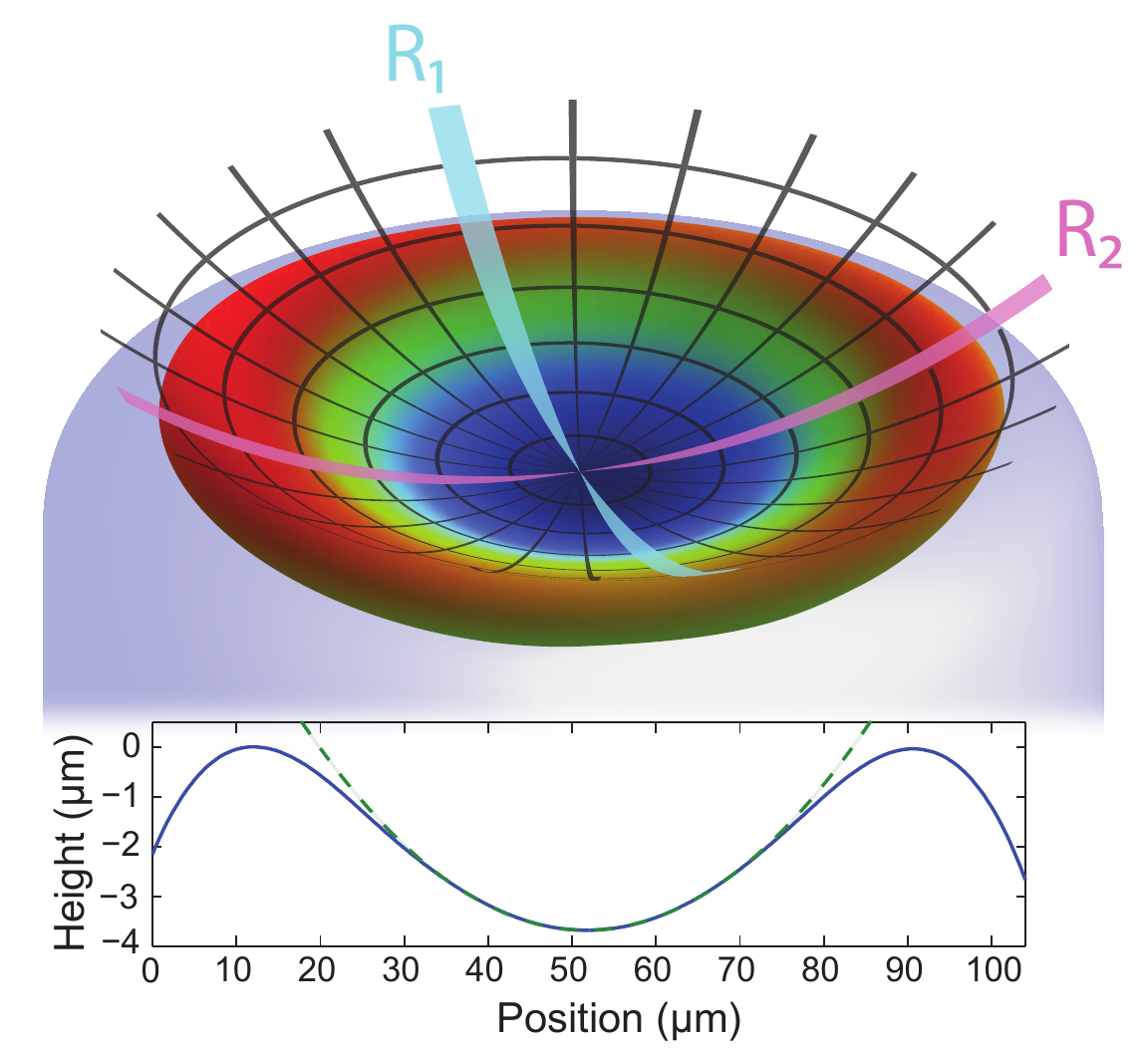}
\caption{\label{fig:FiberendEllipsoid}
CO$_2$ laser-machined fibre end facet imaged with a scanning white-light interferometer (schematic, not to scale). At the centre, the surface can be approximated by an elliptic paraboloid with radii of curvature $R_1$ and $R_2$ along the two principal axes. The inset shows the surface cut along the minor axis (solid blue line) and a fitted parabola (dashed green line).}
\end{figure}

\section{Fabrication of fibre mirrors}
We fabricate the fibre mirror substrates by micromachining end facets of fused silica optical fibres with a CO$_2$ laser \cite{Hunger2012}. Local heating due to the absorption of a laser pulse leads to evaporation of fibre material. In combination with the Gaussian transverse profile of the laser beam, concave structures are formed. Melting should be restricted to a thin layer below the surface of the fibre such that the surface tension leads to an ultrasmooth surface but not to the formation of a convex structure. To minimize the absorption depth, we employ a CO$_2$ laser at a wavelength of 9.3\,\textmu m, close to an absorption maximum of fused silica \cite{Kitamura2007}. Single-mode and multi-mode optical fibres with a cladding diameter of 125\,\textmu m are cleaved, and the end facets are illuminated for 0.65--1.2\,ms with typically 50\,W of laser power focused to a $1/e^2$ beam diameter of 450\,\textmu m. This results in concave structures 73--83\,\textmu m in diameter and with radii of curvature of 120--600\,\textmu m determined by the duration of the laser pulse. The diameter of the structure is limited by the fibre diameter and surface tension, which prevents the formation of sharp edges.

\begin{figure}
\includegraphics[width=1.0\columnwidth]{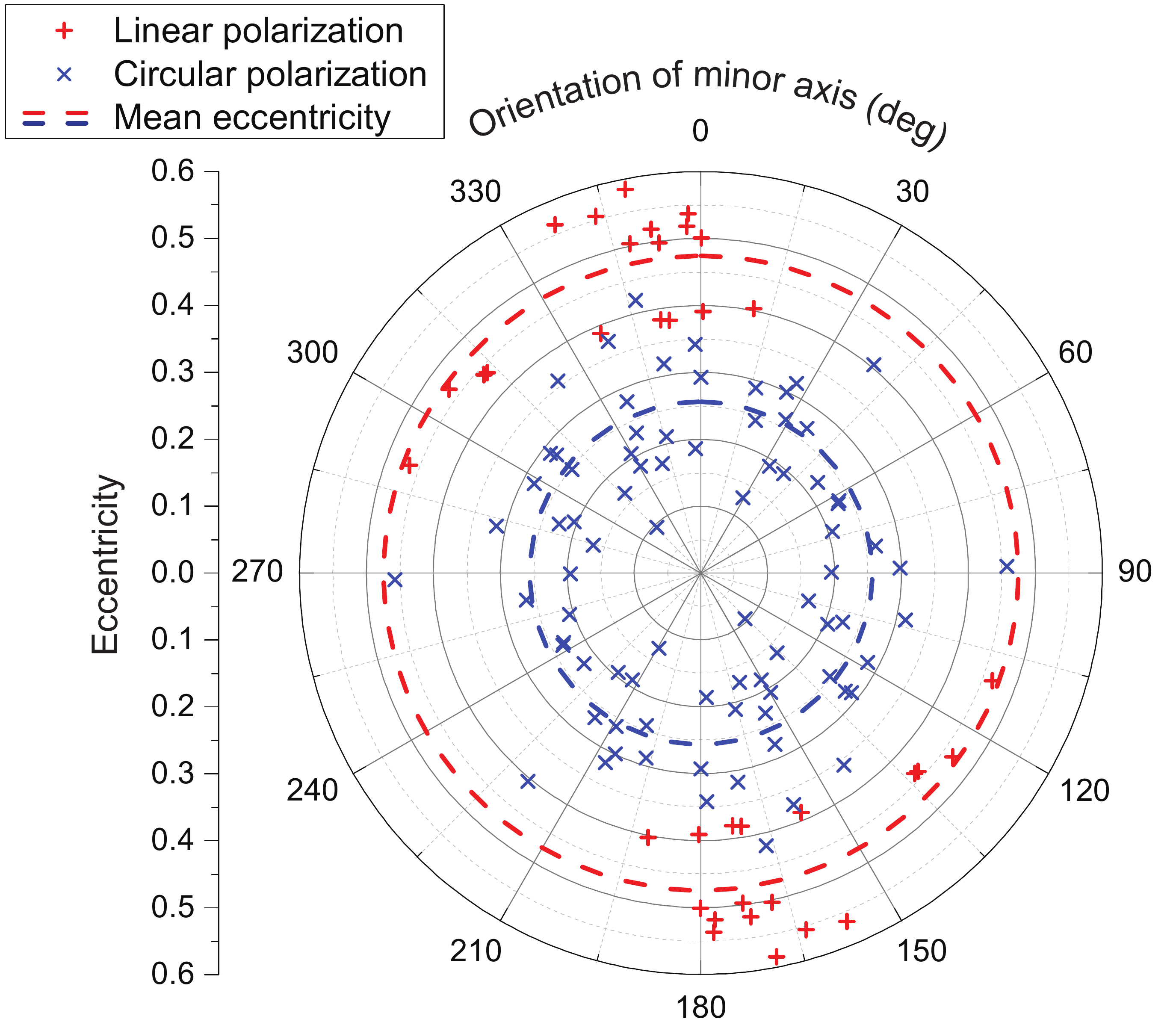}
\caption{\label{fig:EffectLaserPol}
Eccentricity and orientation of the minor axis of the CO$_2$ laser-machined fibre end facets. For samples machined with a linearly polarized laser (red pluses), the orientation is given relative to the polarization axis (vertical in the plot). The mean eccentricity is 0.47 (red circle). Using a circularly polarized laser (blue crosses), the mean eccentricity is reduced to 0.26 (blue circle). To denote the individual axes, the crosses have been duplicated and rotated by 180 degrees, such that each pair of opposing crosses represents one sample. While there is no preferred orientation of the minor axis in the case of circular polarization, the orientations of the minor axes of the fibre surfaces machined with linear polarization are correlated, close to the orientation of the CO$_2$ laser polarization axis.}
\end{figure}

We characterize the fabricated structures with a scanning white-light interferometer. In general, the surface is neither spherical, nor does it exhibit perfect cylindrical symmetry with respect to the fibre axis. Therefore, the radii of curvature are local features and depend on the region of interest when determined from fits to the fibre surface. Near the centre, it is well approximated by an elliptic paraboloid with a major axis having a radius of curvature ($R_1$) larger than that of the minor axis ($R_2$) (figure \ref{fig:FiberendEllipsoid}). We call such mirrors elliptical because the contour lines of their surfaces are ellipses. As can be seen from figure \ref{fig:EffectLaserPol}, we find that the eccentricity $\epsilon=\sqrt{1 - R_2/R_1}$ is influenced by the polarization of the CO$_2$ laser. Using linear polarization, we find a mean eccentricity of 0.47 with a standard deviation of 0.07 over 18 samples. The minor axes of the structures are aligned with the direction of the linear polarization. Switching to circular polarization we find a mean eccentricity of 0.26 with a standard deviation of 0.08 over 41 samples, with no preferred direction for the principal axes (figure \ref{fig:EffectLaserPol}). Apparently, linear polarization of the CO$_2$ laser induces additional asymmetry in the process and leads to an increased eccentricity of the resulting structures.

The machined structures were coated with a commercial high-reflection coating using ion-beam sputtering. Superpolished reference substrates coated in the same run show a transmission of $(2.9 \pm 0.1)$\,ppm at a wavelength of 780\,nm. For a cavity built from two of these reference substrates, we measure a finesse of $340\,000 \pm 20\,000$ using cavity ring-down \cite{Anderson1984}, which corresponds to 6\,ppm additional losses (scattering and absorption) per mirror. Cavities built from two fibre mirrors with the same coating reach a finesse of up to $190\,000 \pm 10\,000$, as determined from direct spectroscopic measurements of the cavity linewidth. This corresponds to parasitic losses of 13.5\,ppm per mirror under the reasonable assumption that the transmission is the same as for the reference substrates.

\section{Theoretical model}\label{sec:Model}

\begin{figure*}
\includegraphics[width=1.8\columnwidth]{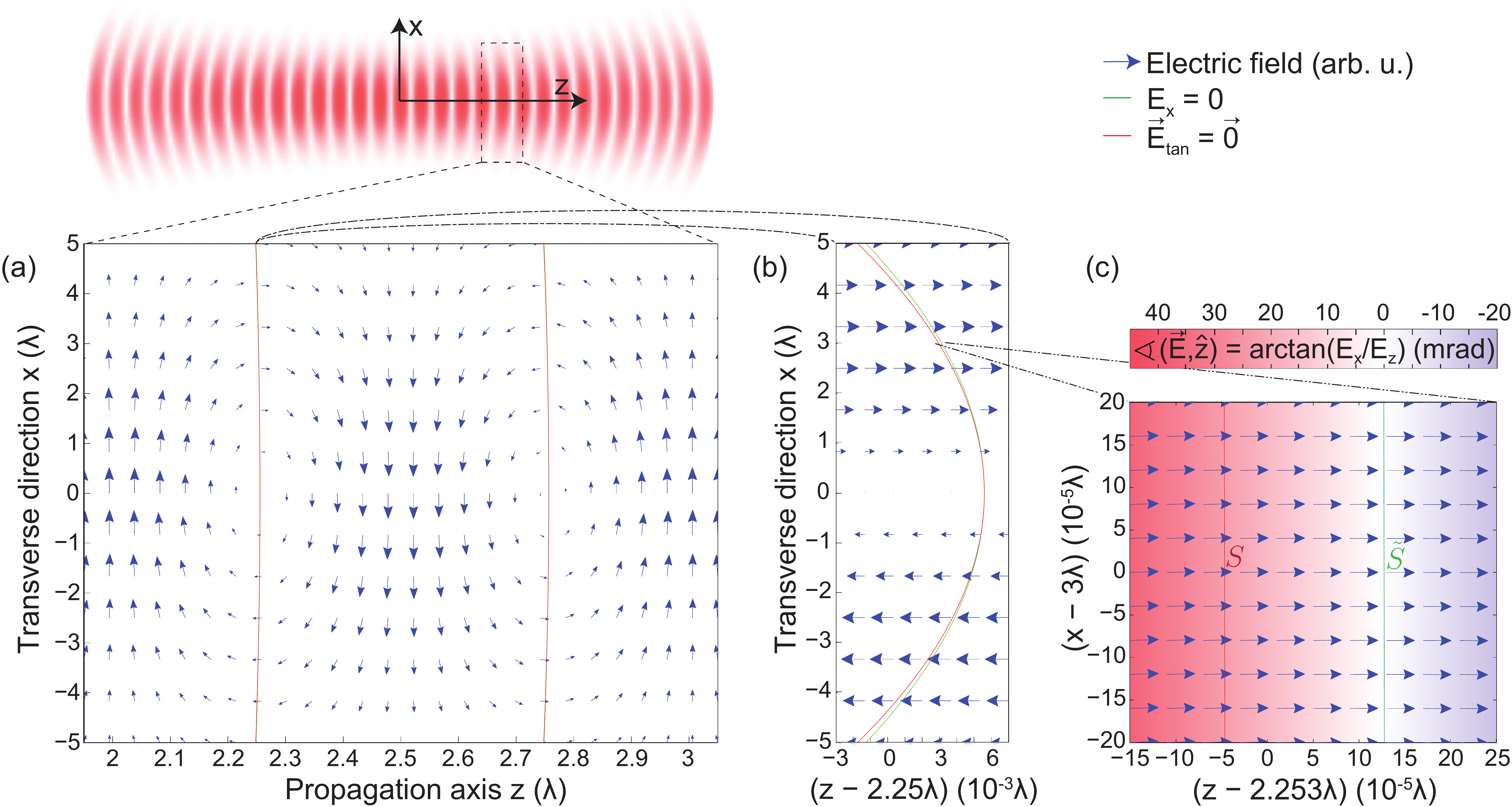}
\caption{\label{fig:IllustrationModel}
Electric field distribution of a fundamental transverse cavity mode mainly polarized along the x-direction with an additional, nonparaxial field component along the propagation axis z. The transverse field component is a Gaussian TEM$_{00}$-mode (wavelength $\lambda=780$\,nm, mode waist 3.5\,\textmu m), from which the longitudinal field component is derived as described in the appendix \cite{Lax1975}. The relative scaling of $E_x$ and $E_z$ corresponds to the scaling of the spatial axes in each subplot. Note that this does not preserve angles [most pronounced in (b)]. Lines indicating the fulfilled conditions $E_x|_{\tilde{S}}=0$ (green) and $\vec{E}_{\text{tan}}|_S=\vec{0}$ (red) are plotted in all three graphs but can only be resolved in (b) and (c). The lines $S$ and $\tilde{S}$ have slightly different radii of curvature and overlap at $x=0$.}
\end{figure*}

In the paraxial resonator theory, the modes of a cavity are described by scalar mode functions, which are solutions of the paraxial wave equation. The resonance frequencies of the modes are deduced from the condition that the field, derived from the mode function, must vanish on the mirror surfaces. Polarization effects do not enter the paraxial resonator theory, which can therefore not account for a frequency splitting of polarization eigenmodes.

Modelling of the frequency splitting of the polarization eigenmodes requires an extension of the scalar theory to a vector theory. Lax \textit{et al.} \cite{Lax1975} pointed out how the scalar paraxial theory is naturally embedded in a more general vector theory. By scaling Maxwell’s equations with the characteristic lengths along the longitudinal and transverse directions of a light beam, they showed that the vector field can be suitably expressed as a power series in the parameter $1/(kw_0)$, which compares the wavelength $\lambda=2\pi/k$ to the beam waist $w_0$. The leading-order term in this series is a transverse field component that satisfies the paraxial wave equation known from the scalar theory, whereas the first-order correction points along the propagation direction of the beam and is out of phase with the main, transverse field component by about 90\textdegree. Such a field is shown in figure \ref{fig:IllustrationModel}, with the transverse field component chosen to point along $x$.

In the case of a vector field, the most natural boundary condition for the electric field on the cavity mirrors is that of a perfect conductor, which imposes that the electric field component tangential to the mirror surface must vanish ($\vec{E}_{\text{tan}}=\vec{0}$). In figure \ref{fig:IllustrationModel}, red lines labelled with $S$ indicate where this condition is fulfilled. A mirror matching one of these lines supports the depicted mode. For comparison of the vector theory with the scalar paraxial theory, green lines labelled with $\tilde{S}$ indicate where the transverse field component, described by a paraxial mode function, vanishes ($E_x=0$). One can see that the green lines, which overlap with the red lines at $x=0$, have a slightly larger radius of curvature [figure \ref{fig:IllustrationModel}(b)]. In the vector treatment of the cavity mode, the transverse part of the field is thus described by a mode function that, at the position of the mirror, has a larger radius of curvature along the polarization direction ($x$) than the mirror itself. The larger radius of curvature of the mode function comes along with a smaller Gouy phase shift and a correspondingly lower resonance frequency than the treatment of the same resonator within the scalar paraxial theory would predict, in which the resonance frequency would be deduced from a mode function with radii of curvature that exactly match the cavity mirrors.

The described frequency correction to the paraxial resonator theory, due to the more accurate treatment of the boundary condition on the mirror surface in a vector theory, was pointed out and calculated by Cullen \cite{Cullen1976} for the fundamental transverse mode of a resonator with spherical mirrors. We consider and calculate this frequency correction for the case of a plano-concave cavity with an elliptical mirror having different radii of curvature $R_1$ and $R_2$ along the major and minor principal axis, respectively. The resulting frequency correction depends only on the radius of curvature along which the mode is polarized. This leads to different frequency corrections $\delta\nu_1$ and $\delta\nu_2$ relative to the paraxial theory for modes which are linearly polarized along the principal axes of the elliptical mirror. The modulus of the (negative) frequency correction is larger for the mode polarized along the minor principal axis, because corrections to the paraxial theory become more important for smaller radii of curvature. Therefore, this mode has the lower resonance frequency.

The consequence of these corrections to the paraxial theory is the qualitatively new effect of a frequency splitting between the polarization eigenmodes of the fundamental transverse mode for a cavity with an elliptical mirror. To the lowest order in $1/(kw_0)$, the frequency splitting is given by (appendix)
\begin{align}
\begin{aligned}
\Delta&\nu = \delta\nu_1-\delta\nu_2 = \frac{\nu_\mathrm{FSR}}{2\pi k}\frac{R_1-R_2}{R_1R_2}.\label{doppelbrech}
\end{aligned}
\end{align}
Here, $\nu_\mathrm{FSR}=c/(2L)$ is the free spectral range of the cavity of length $L$. The frequency splitting of the polarization eigenmodes can be related to the differential phase shift $\Delta\varphi_\mathrm{rt}$ that the two polarization modes acquire during a cavity round trip:
\begin{align}
\Delta\varphi_\mathrm{rt}=\frac{2\pi\,\Delta\nu}{\nu_\mathrm{FSR}}.\label{PFS}
\end{align}
For the considered case of only one elliptical mirror, the differential phase shift per round trip is equal to the differential phase $\Delta\varphi$ the linear polarization modes acquire during reflection from that mirror, $\Delta\varphi_\mathrm{rt}=\Delta\varphi$. This differential phase shift per reflection is therefore given by
\begin{align}
\begin{aligned}
\Delta&\varphi = \frac{1}{k}\frac{R_1-R_2}{R_1R_2} = \frac{\epsilon^2}{kR_2},\label{doppelbrechPhase}
\end{aligned}
\end{align}
which is independent of the cavity length. The relevant geometrical property of the mirror surface in \eqref{doppelbrech} and \eqref{doppelbrechPhase} is the square of the eccentricity, scaled by an additional factor $1/(kR_2) = \lambda / (2\pi R_2)$. The effect of mirror eccentricity thus becomes increasingly important when going for small radii of curvature that approach the size of the wavelength, as, for example, in CO$_2$ laser-machined optical cavities or in the microwave domain. Relation \eqref{doppelbrechPhase} also sets an upper bound of $1/(kR_2)$ for the maximum differential phase shift per reflection for the fundamental transverse mode that can be achieved via mirror asymmetry. It is reached for cylindrical mirrors that have $\epsilon=1$.

\section{Experimental results}\label{sec:HybridCav}

\begin{figure}
\includegraphics[width=1.0\columnwidth]{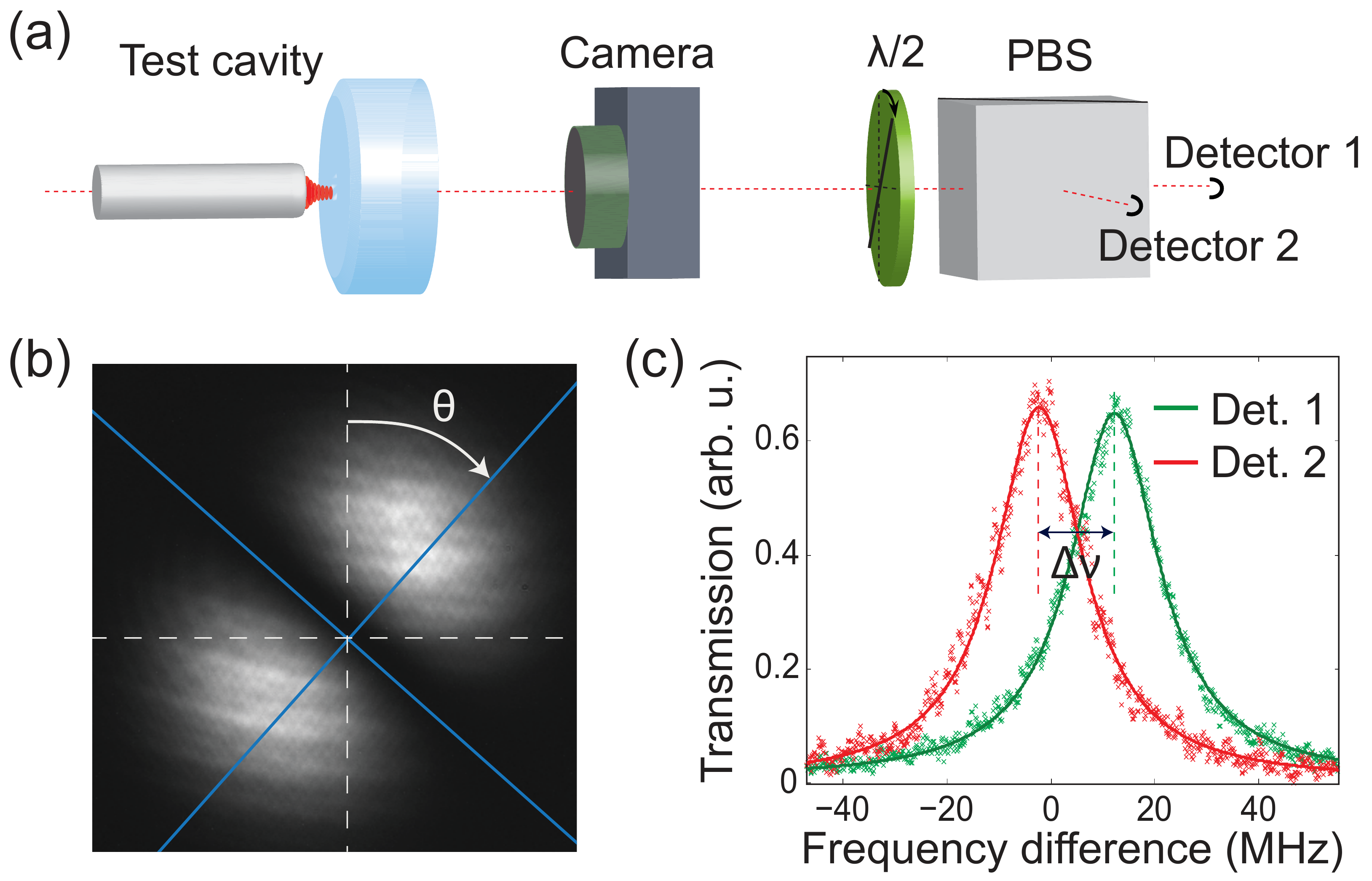}
\caption{\label{fig:HybridCavSetup}
Characterization of the frequency splitting of polarization eigenmodes. (a) Experimental setup. The hybrid cavity consists of one fibre mirror and one macroscopic mirror. (b) Using a removable camera, the transverse mode structure (here TEM$_{01}$) and its orientation $\theta$ can be determined for each cavity resonance. Typical mode waists are 4--7\,\textmu m. (c) The frequency splitting between the two orthogonal, linear polarization eigenmodes is determined from a simultaneous scan over both resonances. The centre of each resonance is determined by fitting Lorentzians (green and red lines) to the measured transmission (green and red crosses).}
\end{figure}

We study the dependence of the frequency splitting of the polarization eigenmodes on the properties of CO$_2$ laser-machined mirrors in a hybrid cavity setup [figure \ref{fig:HybridCavSetup}(a)]. The cavities consist of the fibre mirror under testing and a reference mirror based on a superpolished substrate with a 100\,mm radius of curvature. Typical cavity lengths are around 50\,\textmu m. Using an additional macroscopic mirror and a procedure similar to the one described in \cite{Brandi1997}, the differential phase shift of the reference mirror has been characterized to be smaller than 2\,\textmu rad. This is negligible compared to the phase shift induced by the ellipticity of the fibre mirrors, and we consequently attribute any measured differential phase shift to the fibre mirror under testing.

Light at a wavelength of 780\,nm is coupled into the cavity via the fibre mirror and imaged on a CMOS camera behind the superpolished mirror. The order of the transverse mode and its orientation can thus be assigned to each peak in the transmission spectrum. We observe Hermite-Gaussian modes, as expected for mirrors having an elliptic paraboloid shape [figure \ref{fig:HybridCavSetup}(b)]. From the orientation of the first-order transverse modes and their distance from the fundamental transverse mode in frequency space, the orientation of the major and minor axis of the fibre mirror and the corresponding effective radii of curvature can be deduced \cite{Kogelnik1966}. We optimize the cavity geometry for efficient excitation of the fundamental transverse mode and only measure at cavity lengths where no hints of higher-order transverse modes are visible simultaneously with the modes to be measured.

\begin{figure}
\includegraphics[width=1.0\columnwidth]{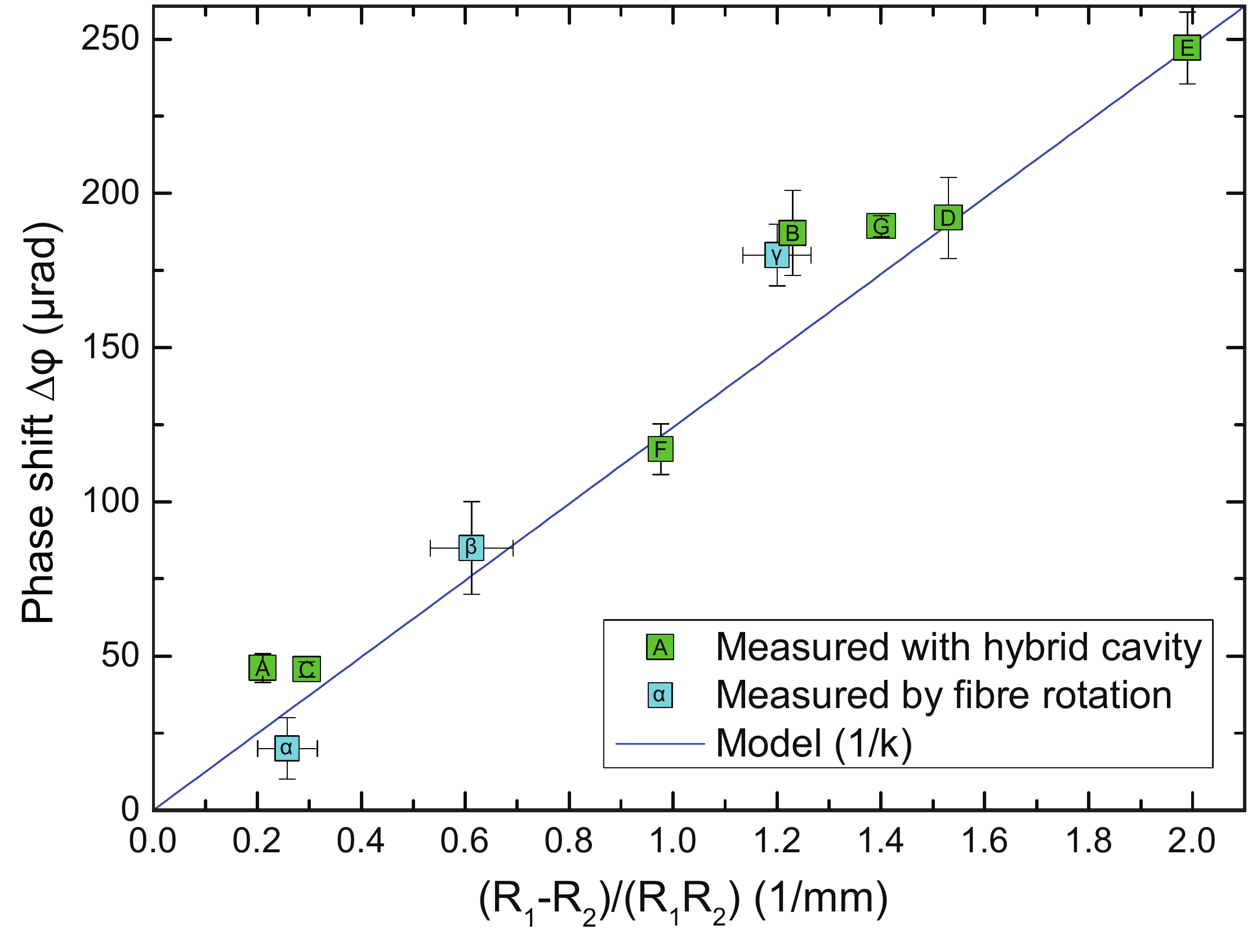}
\caption{\label{fig:BirefVsEllipt}
Differential phase shift per reflection for fibre mirrors as a function of the mirror geometry. Green squares: Phase shift measured with a hybrid cavity, radii of curvature determined from the transverse mode distances. Error bars are statistical. Cyan squares: Phase shift measured by rotation using a fibre cavity (section \ref{sec:Rotation}), radii of curvature determined from surface fits. Error bars are confidence intervals deduced from the fits. The letters indicate different fibre mirrors. The blue line with slope $1/k$ is the result of a theoretical model (section \ref{sec:Model}) with no free parameter.}
\end{figure}

To characterize the polarization eigenmodes of the cavity, we replace the camera with a $\lambda/2$-waveplate, a polarizing beam splitter, and a photomultiplier tube at each output port of the beam splitter [figure \ref{fig:HybridCavSetup}(a)]. The polarization of the incoming light is adjusted to equally excite both polarization eigenmodes. The waveplate in the detection setup is adjusted until the light of each polarization eigenmode is mapped onto one detector [figure \ref{fig:HybridCavSetup}(c)]. This is possible because the polarization eigenmodes are linear within the measurement accuracy. For every fibre mirror characterized in this way, we have verified that the polarization eigenmodes are aligned with the principal axes of the CO$_2$ laser-machined structure, with the mode assigned to the minor axis always being lower in frequency. The frequency splitting $\Delta\nu$ of the two polarization eigenmodes is measured by scanning the cavity length with a linear ramp while phase modulating the probe light with an electro-optic modulator to generate sidebands as frequency markers. To compensate for a potential difference in signal path delays, we switch scanning directions and take the mean of 100 scans in each direction. The cavity length is determined from a measurement of the free spectral range $\nu_\mathrm{FSR}$ using two lasers which are simultaneously resonant with neighbouring fundamental transverse cavity modes. We can then convert the cavity length-dependent $\Delta\nu$ into the differential phase shift per cavity round trip [equation (\ref{PFS})], which is independent of the cavity length. It equals the differential phase shift per reflection of the fibre mirror if the influence of the superpolished mirror is negligible. The results as a function of the radii of curvature are shown in figure \ref{fig:BirefVsEllipt}, showing excellent agreement with the theoretical model. This demonstrates that ellipticity of the cavity mirrors is the dominant reason for the splitting of polarization modes in cavities built from the presented CO$_2$ laser-machined mirrors.

An alternative way to determine the radii of curvature, instead of measuring the frequency separation of transverse modes, is to fit an elliptic paraboloid to the measured surface profile. The fit is weighted with the estimated transverse profile of the cavity mode, because deviations from a spherical surface lead to a spatial dependence of the curvature. The resulting radii of curvature, and thus the resulting differential phase shift per reflection, depend on the position of the cavity mode on the mirror. For mirrors based on single-mode fibres, the core of the fibre can be used as a position reference when the overlap of cavity mode and fibre mode is optimized by maximizing the transmission of the cavity. For mirrors based on single-mode fibres, we find good agreement between the values obtained from the fit and those from the transverse mode distances. It is therefore possible to predict the geometrically induced differential phase shift of a particular machined structure directly from a measurement of its eccentricity and radius of curvature without the need to build a cavity or to apply a reflective coating.

\section{Control by fibre rotation}\label{sec:Rotation}
\begin{figure}
\includegraphics[width=1.0\columnwidth]{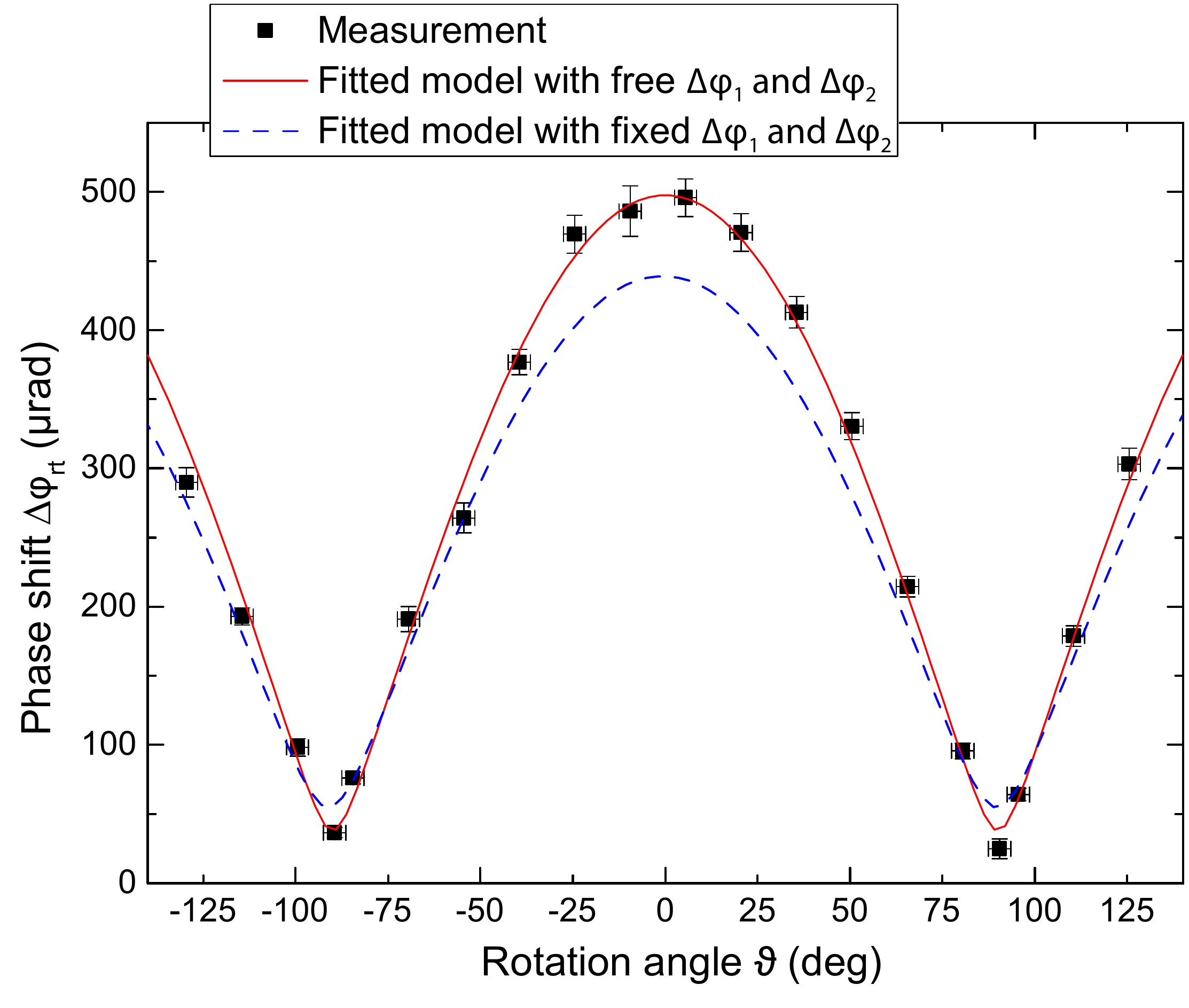}
\caption{\label{fig:BirefRotation}
Dependence of the frequency splitting of polarization eigenmodes of the fundamental transverse mode on the relative rotation angle of the two elliptical cavity mirrors. The error bars denote the statistical standard error for the phase shift and the estimated uncertainty for the rotation angle. See text for a description of the model used to fit the data.}
\end{figure}

The rotational alignment of two elliptical cavity mirrors, i.e., the relative orientation of their major axes, can be used to further control the frequency splitting of the polarization modes \cite{Jacob1995}. Figure \ref{fig:BirefRotation} shows the results of a measurement of the differential phase shift per round trip of a fibre cavity as a function of the angle by which one of the fibre mirrors was  rotated around the fibre axis. The fibre cavity was made of two single-mode fibres. A polarization-resolving excitation and detection setup was used, analogous to the one described in section \ref{sec:HybridCav}. Careful optimization of the cavity transmission was found to be crucial for reproducible measurements.

We use a model based on the Jones formalism to describe the measured data \cite{Moriwaki1997,Brandi1997}. In this model, the two cavity mirrors are characterized by their individual differential phase shifts per reflection $\Delta\varphi_1$ and $\Delta\varphi_2$. They can be rotated around the cavity axis with the relative angle between their major axes denoted by $\vartheta$. The differential phase shift per round trip $\Delta\varphi_\mathrm{rt}$ of the two polarization eigenmodes of the cavity is deduced from the eigenvalues of the Jones matrix describing a round trip of a polarization vector through the cavity. It is given by
\begin{align}
\Delta\varphi_\mathrm{rt}=\sqrt{\Delta\varphi_1^2+\Delta\varphi_2^2+2\Delta\varphi_1\Delta\varphi_2\cos\left(2\vartheta\right)}.\label{rotation}
\end{align}

For the fits in figure \ref{fig:BirefRotation}, an offset angle $\vartheta_0$ was introduced as an additional fit parameter to account for the unknown initial orientation of the fibres. The fibres used for the depicted measurement are the ones that yielded data points D and E in figure \ref{fig:BirefVsEllipt}, with measured differential phase shifts $\Delta\varphi_1 = (192 \pm 13)$\,\textmu rad and $\Delta\varphi_2 = (247 \pm 12)$\,\textmu rad, respectively. The dashed blue curve in figure \ref{fig:BirefRotation} is a fit of (\ref{rotation}) using these measured values for $\Delta\varphi_1$ and $\Delta\varphi_2$, such that $\vartheta_0$ is the only free parameter. The red solid curve is a fit with the differential phase shifts of the two mirrors as free parameters, resulting in $\Delta\varphi_1 = (230 \pm 11)$\,\textmu rad and $\Delta\varphi_2 = (268 \pm 9)$\,\textmu rad.  The discrepancy in the results of the two methods can be attributed to different positions of the cavity mode on the fibre mirrors. Using weighted surface fits as described above, we have studied the local eccentricity of the mirror surface as a function of lateral displacement. We find that the eccentricity can change significantly for lateral displacements of only a few micrometres.

An obvious but important implication of (\ref{rotation}) is that the frequency splitting of polarization eigenmodes of a cavity can be tuned via the rotation angle $\vartheta$. This is experimentally confirmed by the data in figure \ref{fig:BirefRotation}. When the major axes of the mirrors are parallel, the differential phase shift per cavity round trip, and thus the frequency splitting of the polarization modes, is maximized with $\Delta\varphi_\mathrm{rt}^{\text{max}}=\Delta\varphi_1+\Delta\varphi_2$. A minimal value of $\Delta\varphi_\mathrm{rt}^{\text{min}}={|\Delta\varphi_1-\Delta\varphi_2|}$ is achieved when the major axes of the two mirrors are perpendicular. In particular, the frequency splitting of the polarization eigenmodes vanishes in this configuration if the differential phase shifts per reflection of the two mirrors are equal ($\Delta\varphi_1 = \Delta\varphi_2$). This is to be expected, because the effects of the two identical mirrors with perpendicular rotational orientation counterbalance each other.

\section{Conclusion}
We have identified elliptical mirror surfaces as the dominant source of the frequency splitting of the polarization eigenmodes of the fundamental transverse resonator mode in CO$_2$ laser-machined optical cavities. Using a CO$_2$ laser at a wavelength of 9.3\,\textmu m, we fabricated structures with ultrasmooth surfaces on fibre end facets. In combination with a highly reflective coating, finesses of up to 190\,000 were reached, allowing for a high spectral resolution in our measurements. We find excellent qualitative and quantitative agreement between experimental data and a mathematical model based on corrections to the paraxial resonator theory \cite{Cullen1976}, which relies only on the mirror geometry. The agreement includes the orientation of the polarization eigenmodes along the principal axes of the elliptical mirror, the fact that the polarization mode which is polarized along the minor axis has the lower resonance frequency, and the magnitude of the frequency splitting.

We have identified the shape of the fabricated mirror surfaces as the crucial control parameter for the splitting of polarization eigenmodes. This suggests that the final splitting can to a large degree be controlled during the CO$_2$ laser-machining process. Consequently, optimization is possible without the necessity to build cavities or even coat the fibre end facets. An excellent tuning mechanism for the frequency splitting of cavities built from existing mirrors is given by the rotation of one of the mirrors around the cavity axis. The relative angle between the major axes of the two mirrors can be used to alter and, in the case of mirrors with identical differential phase shifts, even to cancel the frequency splitting. Alternatively, the frequency splitting can be set to nonvanishing values in a controlled manner by fabricating suitable asymmetric mirrors.

There are a number of quantum information protocols, for example entanglement generation \cite{Wilk2007} or quantum state transfer in a cavity-based quantum network \cite{Ritter2012}, which rely on degenerate polarization modes. The concepts and methods presented here should allow for future experiments implementing these protocols using high-finesse fibre-based optical Fabry-Perot resonators with degenerate polarization eigenmodes.

\begin{acknowledgments}
We thank J. Lang and R. Schittko for their contributions during the early stages of the experiment, H. Specht for AFM measurements, and J. Reichel, D. Hunger and J. Hare for helpful discussions concerning the fabrication of CO$_2$ laser-machined mirrors. This work was supported by the DFG via NIM and by the Bundesministerium für Bildung und Forschung via IKT 2020 (Q.com).
\end{acknowledgments}

\setcounter{equation}{0}
\renewcommand{\theequation}{A.\arabic{equation}}
\section*{Appendix: Calculation of the frequency splitting of polarization eigenmodes for a cavity with one elliptical mirror}
The calculation behind \eqref{doppelbrech} follows a publication by Cullen \cite{Cullen1976}. The results are extended to the case of cavities with one elliptical mirror and to arbitrary Hermite-Gaussian modes. Elliptical cavity mirrors are essential for obtaining the frequency splitting of the polarization eigenmodes of the fundamental transverse mode we observe.

The calculation is based on Green's second identity, which states that for two functions $f$ and $g$, which are twice continuously differentiable the relation
\begin{equation}
\int_{V}\left(f\triangle g-g\triangle f\right)\mathrm{dV}=\int_{S}\left(f\vec{\nabla}g-g\vec{\nabla}f\right)\cdot\vec{\mathrm{dS}}\label{GreensIdent}
\end{equation}
holds. Here, $V$ is a volume with a surface $S$, of which $\vec{\mathrm{dS}}$ is an outward-pointing infinitesimal area element \cite{Erickson1975}.
We consider the case that $f\equiv E_x$ is a mode function, which we assume to represent the x-component of the electric field of a resonator mode. We assume that $E_x$ fulfils the Helmholtz equation, $\triangle E_{x}+k^{2}E_{x}=0$, and require it to vanish on the mirror surface, $E_{x}|_{S}=0$, according to the boundary condition for mode functions in the paraxial resonator theory. The wavenumber $k$ is directly related to the frequency $\nu =kc/ (2\pi)$ of the mode.
The function $g\equiv \tilde{E}_{x}$ is given by an almost identical mode function $\tilde{E}_{x}\cong E_{x}$, which describes the transverse field component of a vector mode, as discussed in section \ref{sec:Model}. It does not exactly vanish on the mirror surface but takes on nonzero values $\tilde{E}_{x}^{S}\equiv\tilde{E}_{x}|_{S}$, which fulfil the boundary condition of a perfect conductor $\vec{E}_{\text{tan}}|_{S}=\vec{0}$, where $\vec{E}_{\text{tan}}$ designates the electric field component, which is tangential to the mirror surface. $\tilde{E}_{x}$ has a slightly different frequency $\tilde{\nu}$ and is assumed to fulfil the Helmholtz equation $\triangle\tilde{E}_{x}+\tilde{k}^{2}\tilde{E}_{x}=0$.
Inserting these conditions in \eqref{GreensIdent} and assuming that the frequency shift between the two modes, $\delta\nu=\tilde{\nu}-\nu$, is small ($|\delta\nu|\ll\nu$), leads to the following equation for the frequency shift \cite{Cullen1976}:

\begin{equation}
\delta\nu\approx\frac{c}{4\pi k}\,\frac{\int_{S}\tilde{E}_{x}^{S}\vec{\nabla}E_{x}\cdot\vec{\mathrm{dS}}}{\int_{V}E_{x}^{2}\;\mathrm{dV}}\label{integralDeltanu}.
\end{equation}
Here, the volume integral has to be taken over the whole resonator volume and the surface integral in general over both resonator mirrors. In order to specify the amount of frequency splitting of the polarization eigenmodes induced by a single elliptical cavity mirror, we consider a plano-concave cavity. In this configuration, $\tilde{E}_{x}^{S}=0$  on the planar mirror. Therefore, the surface integral in \eqref{integralDeltanu} can be restricted to the surface of the concave mirror. On the left-hand side of \eqref{integralDeltanu}, terms of the relative order $\mathcal{O}\left(\delta\nu/\nu\right)$ have been neglected.

We take $E_{x}$ to be a Gaussian TEM$_{nm}$-mode. These modes fulfil the paraxial wave equation and not the Helmholtz equation, as was assumed above. The approximation made in going from the Helmholtz equation to the paraxial wave equation is, however, polarization independent as long as the mode functions are polarization independent. Any error resulting from this approximation will thus not affect the frequency splitting of polarization eigenmodes in a first-order perturbation calculation. In complex notation, the mode functions are given by
\begin{align}
&E^{\mathbb{C}}_{x}(x,y,z)=e^{ikz}\,u_{n}^{x}(x,z)\,u_{m}^{y}(y,z),\label{gaussMode}\\
&u_{n}^{x}(x,z)=\frac{\sqrt[4]{\frac{2}{\pi}}}{\sqrt{2^nn!w_{x}(z)}}\,H_n\left(\frac{\sqrt{2}x}{w_x(z)}\right)\notag\\&\qquad\times\exp\left(-i\left(n+\frac{1}{2}\right)\xi_{x}(z)+i\frac{kx^{2}}{2R_{x}(z)}-\frac{x^{2}}{w_{x}(z)^{2}}\right),\notag
\end{align}
where $w_x(z)$ is the mode field radius along x, $R_x(z)$ is the radius of curvature of the wavefronts along x, $\xi_x(z)$ is the corresponding Gouy phase, and $H_n$ denotes a Hermite polynomial of degree $n$. $R_x(z)$, $w_x(z)$ and $\xi_x(z)$ are functions of the mode waist $w_{0x}$, $k$, and $z$ and follow the usual definitions \cite{Kogelnik1966}. The expressions for  $u_{m}^{y}(y,z)$ are analogous. We explicitly consider the case of a resonator with an elliptical mirror $(R_{x}\neq R_{y})$, meaning that the mode waists $w_{0x}$ and $w_{0y}$ and all  derived parameters are different for $u_{n}^{x}(x,z)$ and $u_{m}^{y}(y,z)$. The z-dependence of the parameters $w_{x}(z)$, $R_{x}(z)$, and $\xi_{x}(z)$ will from now on be omitted in the notation.

We first consider the surface integral in \eqref{integralDeltanu}. To calculate the function $\tilde{E}_{x}^{S}$, we further follow the procedure used by Cullen \cite{Cullen1976}: starting from $E_x$, we will calculate the longitudinal field component $E_{z}$ and will use its value $E_{z}^{S}$ on the mirror surface to derive an expression for $\tilde{E}_{x}^{S}$ which satisfies the boundary condition $\vec{E}_\mathrm{tan}|_{S}=\vec{0}$.

To calculate the longitudinal field component $E_z$, one can follow Cullen \cite{Cullen1976} or use the more rigorous results of Lax \textit{et al.}  \cite{Lax1975}. The electric field of a resonator mode written as a power series in $\zeta \equiv 1/(kw_0)$ fulfils Maxwell's equations if the zeroth-order term is a transversal field component (here $E_x$)  that satisfies the paraxial wave equation and the first-order correction is a longitudinal field component (here $E_z$) given by
\begin{equation}
E^{\mathbb{C}}_z=\frac{i}{k}\frac{\partial E^{\mathbb{C}}_x}{\partial x}. \label{ezKomplex}
\end{equation}
Further corrections to  $E_z$ are of the order $\mathcal{O}(\zeta^3)$; i.e., $E^{\mathbb{C}}_z=\frac{i}{k}\frac{\partial E^{\mathbb{C}}_x}{\partial x}(1+ \mathcal{O}(\zeta^2))$.

We consider standing-wave solutions in the form of the imaginary part of the complex mode functions,
\begin{align}
E_x=\operatorname{Im}\left(E^{\mathbb{C}}_x\right),\ \ \ E_z=\operatorname{Im}\left(E^{\mathbb{C}}_z\right),\label{exStanding}
\end{align}
corresponding to modes which have a node at the planar cavity mirror placed at $z=0$.
Close to the centre of the concave mirror with coordinates $(0;0;L)$, $L>0$, the mirror surface $S$ is in parabolic approximation given by
\begin{equation}
S:\quad z=L-\frac{x^{2}}{2R_{x}}-\frac{y^{2}}{2R_{y}}.\label{mirrorSurf}
\end{equation}
Equations \eqref{ezKomplex}, \eqref{exStanding}, and \eqref{mirrorSurf} and the resonance condition $E_{x}|_{S}=0$ yield an expression for the field component $E_{z}^{S}$ on the mirror surface as a function of $x$ and $y$.
This field component $E_{z}^{S}$ varies only slowly close to the mirror surface and is used to deduce $\tilde{E}_{x}^{S}$. The boundary condition  $\vec{E}_\mathrm{tan}=\vec{0}$, which $\tilde{E}_{x}^{S}$ and $E_{z}^{S}$ are supposed to fulfil,  is equivalent to the condition that the electric field $\vec{E}$ is parallel to the normal $\vec{N}$ to the mirror surface $S$. Based on \eqref{mirrorSurf}, the normal can be given by $\vec{N}=\left(x/R_{x};y/R_{y};1\right)^{\mathrm {T}}$ (not normalized).
Using $\vec{E}\parallel\vec{N}$ and \eqref{ezKomplex}, \eqref{exStanding}, and \eqref{mirrorSurf}, we thus get the first intermediate result:
\begin{equation}
\tilde{E}_{x}^{S}=\frac{x}{R_{x}}E_{z}^{S}=\frac{x}{kR_{x}}\operatorname{Re}\left(\left.\frac{\partial E^{\mathbb{C}}_x}{\partial x}\right|_{S}\right).\label{AppexTildeSurf}
\end{equation}

The remaining constituent for the surface integral in \eqref{integralDeltanu} is the inner product $\vec{\nabla}E_{x}\cdot\vec{\mathrm{dS}}$ on the mirror surface $S$.
The outward-pointing infinitesimal surface element $\vec{\mathrm{dS}}$ is deduced from the gradient to the mirror surface \eqref{mirrorSurf} and is given by $\vec{\mathrm{dS}}=\left(x/R_{x};y/R_{y};1\right)^{\mathrm{T}}\;\mathrm{dx}\,\mathrm{dy}$. The gradient $\vec{\nabla}E_{x}$ is dominated by the z-dependence of the carrier. Relative to this contribution, additional contributions from the envelope are suppressed by a factor of $\mathcal{O}(\zeta^2)$ for the z-direction and $\mathcal{O}(\zeta)$ for the transverse directions.  Due to the additional factors $x/R_x$  and $y/R_y$ from the inner product with $\vec{\mathrm{dS}}$, the corrections due to the transverse contributions will be further suppressed to the relative order $\mathcal{O}(\zeta^2)$  in the final surface integral \eqref{AppoberflIntegral}. Neglecting these higher-order contributions, we find on the surface $S$:
\begin{align}
\begin{aligned}
&\vec{\nabla}E_{x}\cdot\vec{\mathrm{dS}}\approx\frac{\partial E_{x}}{\partial_z}\;\mathrm{dx}\,\mathrm{dy}\\&\approx\operatorname{Im}(ikE^{\mathbb{C}}_x)\;\mathrm{dx}\,\mathrm{dy}=-k\operatorname{Re}(E^{\mathbb{C}}_x)\;\mathrm{dx}\,\mathrm{dy}.
\end{aligned}\label{AppgradCdotOberflel}
\end{align}

Using \eqref{AppexTildeSurf}, \eqref{AppgradCdotOberflel}, and the boundary condition $E_{x}|_{S}=0$, integration by parts leads to an expression for the surface integral in \eqref{integralDeltanu}:
\begin{align}
\begin{aligned}
&\int_S\tilde{E}^S_x\,\vec{\nabla}E_x\cdot\vec{\text{dS}}\\&=\frac{1}{R_x}\iint x\operatorname{Re}(E^{\mathbb{C}}_x)\,\frac{\partial \operatorname{Re}(E^{\mathbb{C}}_x)}{\partial x}\;\text{dx}\,\text{dy}=-\frac{1}{2R_x},\label{AppoberflIntegral}
\end{aligned}
\end{align}
which is in principle independent of the transverse mode chosen for $E^{\mathbb{C}}_x$. Corrections due to the approximations made above are of the relative order $\mathcal{O}(\zeta^2)$.

The denominator of \eqref{integralDeltanu} is calculated from the real version of the fields, i.e., \eqref{exStanding}. As  the complex mode function \eqref{gaussMode} is normalized, integration over the real field yields
\begin{equation}
\int_{V}E_{x}^{2}\;\mathrm{dV}=\frac{L}{2}.\label{AppvolumeInt}
\end{equation}
For the TEM$_{00}$-mode, corrections to this integral due to the replacement $\tilde{E}_xE_x\rightarrow E_x^2$ in the derivation of \eqref{integralDeltanu} are of the relative order $\mathcal{O}(\zeta^2)$.

By combining \eqref{integralDeltanu}, \eqref{AppoberflIntegral}, and \eqref{AppvolumeInt}, we get the frequency correction for the considered vectorial mode relative to the prediction of the paraxial theory, due to the refined boundary condition $E_{x}|_S=0\rightarrow\vec{E}_\mathrm{tan}|_S=\vec{0}$,
\begin{equation}
\delta\nu_{\text{PolX}}\approx-\frac{c}{4\pi kL}\frac{1}{R_x},\label{frequenzkorrektur}
\end{equation}
up to terms of the relative order $\mathcal{O}(\zeta^2)$.

The frequency correction $\delta\nu_{\text{PolY}}$ for the corresponding mode which is quasi-linearly polarized along the y-direction is analogous with $R_x$ replaced by $R_y$. The difference between these terms is the frequency splitting of the two polarization eigenmodes:
\begin{align}
\begin{aligned}
\Delta\nu = \delta\nu_\mathrm{PolX}-\delta\nu_\mathrm{PolY} = - \frac{c}{4\pi kL}\frac{R_y-R_x}{R_yR_x}.\label{Appdoppelbrech}
\end{aligned}
\end{align}
The frequency splitting is negative (positive) when $R_y>R_x$ ($R_y<R_x$), meaning that the polarization mode that is polarized along the smaller radius of curvature has the lower frequency.

Going from \eqref{frequenzkorrektur} to \eqref{Appdoppelbrech} requires the frequency splitting $\Delta\nu$ to be large compared to the terms which have been neglected in the calculation of the individual frequency shifts $\delta\nu$. This puts a limit on the validity of \eqref{Appdoppelbrech}, which requires $\Delta\nu\gg  \zeta^4\nu_{\text{FSR}}/(2\pi)$ or $\Delta\varphi_\mathrm{rt}\gg\zeta^4$ [see equation \eqref{PFS}]. For the cavities characterized in this paper ($\zeta\lessapprox1/30$), this requirement corresponds to $\Delta\varphi_\mathrm{rt}\gg 1$\,\textmu rad, which is clearly fulfilled for all measured cavities.

\end{document}